\documentclass[draftclsnofoot,onecolumn,12pt,twoside]{IEEEtran}

\usepackage[utf8]{inputenc} 
\usepackage[T1]{fontenc}
\usepackage{url}              
\usepackage{cite}             

\usepackage[cmex10]{amsmath}  
\usepackage{mleftright}       
\mleftright                   

\usepackage{graphicx}         
\usepackage{booktabs}         

\usepackage{amssymb}
\usepackage{amsthm}
\usepackage{amsfonts}
\usepackage{tcolorbox}

\makeatletter
\renewcommand*\env@matrix[1][\arraystretch]{%
  \edef\arraystretch{#1}%
  \hskip -\arraycolsep
  \let\@ifnextchar\new@ifnextchar
  \array{*\c@MaxMatrixCols c}}
\makeatother

\newcommand{\B}{\boldsymbol}

\newcommand{\F}{\mathbb{F}}

\newcommand{\vand}{\mathsf{Vand}}

\newtheorem{proposition}{Proposition}
\newtheorem{lemma}{Lemma}
\newtheorem{observation}{Observation}
\newtheorem{corollary}{Corollary}
\newtheorem{remark}{Remark}
\newtheorem{claim}{Claim}
\newtheorem{definition}{Definition}
\newtheorem{theorem}{Theorem}

\title{On the Structure of 
Higher Order MDS Codes} 
\author{Harshithanjani Athi, Rasagna Chigullapally, Prasad Krishnan and Lalitha Vadlamani
\thanks{\hrule}%

\thanks{Harshithanjani Athi, Rasagna Chigullapally, Dr. Krishnan and Dr. Lalitha are with the Signal Processing and Communications Research Center, International Institute of Information Technology, Hyderabad, 500032, India (email: $\{$harshithanjani.athi@research., rasagna.c@research., prasad.krishnan@,lalitha.v$\}$iiit.ac.in). Acknowledgments: Harshithanjani and Rasagna are supported by the 2022 Qualcomm Innovation Fellowship (India).  Dr. Krishnan acknowledges support from SERB-DST project CRG/2019/005572. }}

\begin{document}

\maketitle

\begin{abstract}
 A code of length $n$ is said to be (combinatorially) $(\rho,L)$-list decodable if the Hamming ball of radius $\rho n$ around any vector in the ambient space does not contain more than $L$ codewords.  We study a recently introduced class of higher order MDS codes, which are closely related (via duality) to codes that achieve a generalized Singleton bound for list decodability. For some $\ell\geq 1$, higher order MDS codes of length $n$, dimension $k$, and order $\ell$ are denoted as $(n,k)$-MDS($\ell$) codes. We present a number of results on the structure of these codes, identifying the `extend-ability' of their parameters in various scenarios. Specifically, for some parameter regimes, we identify conditions under which $(n_1,k_1)$-MDS($\ell_1$) codes can be obtained from $(n_2,k_2)$-MDS($\ell_2$) codes, via various techniques.  We believe that these results will aid in efficient constructions of higher order MDS codes. We also obtain a new field size upper bound for the existence of such codes, which arguably improves over the best known existing bound, in some parameter regimes. 
\end{abstract}


\section{Introduction}
\label{sec:introduction}
The notion of list decoding was introduced by \cite{Eli57} as a generalization of unique decoding and helps in handling greater number of errors than that allowed by unique decoding. List decoding is an important tool in capacity achieving codes, in a variety of channels, with applications in various fields like complexity theory and cryptography.

List decoding of Reed-Solomon (RS) codes is of specific interest, since RS codes are known to be Maximum Distance Separable (MDS) codes, i.e., they have the largest possible rate for a given minimum distance. A celebrated work of Guruswami and Sudan \cite{GSU99} showed that one can efficiently list decode RS codes (via a Berlekamp-Welch type polynomial interpolation algorithm) up to the so-called Johnson radius \cite{John62}, an upper bound on the decoding radius for which polynomial list sizes are guaranteed. Continuing this line of research, it was shown in \cite{RudraWooters} that there exists a large class of RS codes which are  list decodable (that is, have small list sizes) beyond the Johnson bound as well. 

Deviating from previous approaches of obtaining efficient algorithms for list decoding, recent work (for instance, \cite{Sto20,Rot22,GST22}) has focused on the question of \textit{combinatorial} list decodability of codes, especially MDS (and RS) codes. 

\textit{{Notation:}} We let $\F$ be the finite field with $q$ elements (we suppress the field size unless required explicitly). The notation $[n]$ denotes $\{1,\hdots,n\}$. The binomial coefficient with parameters $m,r$ is denoted by $\binom{m}{r}$. A linear $(n,k)$-code $\mathcal{C}$ over $\F$ is a $k$-dimensional subspace of $\F^n$. We denote the dual code of $\mathcal{C}$ by $\mathcal{C}^\perp$ which is an $(n,n-k)$-code. The zero-vector is denoted by $\B 0$, which we also slightly abuse for denoting the zero-dimensional subspace of a vector space.  For a matrix $V$ with $n$ columns over a field and subset $A \subseteq [n]$, we denote by $V_{A}$ the span of columns of $V$ indexed by $A$. For convenience, we also abuse this notation $V_A$ slightly, to occasionally denote  the submatrix of $V$ consisting of the columns in $A$ (the exact meaning of $V_A$ should be clear from the context). For set $A$, we let $\binom{A}{j}$ denote the set of all subsets of $A$ of size $j$.  Without loss of generality is abbreviated as WLOG.

Formally, a \textit{(combinatorial) list decodable} code is defined as follows:
\begin{definition}
    A code $\mathcal{C}$ of length $n$ over $\mathbb{F}$ is said to be $(\rho,L)$-list decodable if for every ball of radius $\rho n$ around any vector in $\mathbb{F}^n$, there are at most $L$ codewords of $\mathcal{C}$.
\end{definition} 
For $L=1$, the maximum decoding radius is given by the Singleton bound:
\begin{equation*}
    \rho \leq \frac{1-R}{2}.
\end{equation*}
Codes for which the Singleton bound is met with equality are called Maximum Distance Separable (MDS) codes. 

A generalization of the Singleton bound was recently proved for list-decoding in \cite{Sto20,Rot22,GST22}.
\begin{proposition}{(Generalized Singleton bound)}
    If an $(n,k)$-code $\cal C$ over an alphabet of size $q$ is $(\rho, L)$-list-decodable, then 
\begin{equation*}
    |\mathcal{C}| \leq L q^{ n - \lfloor \frac{(L+1)\rho n}{L} \rfloor}.
\end{equation*}
\end{proposition}
The following definition captions the notion of \textit{average-radius list decodability}. 
\begin{definition}\cite{VN14}
    A code $\mathcal{C}$ of length $n$ over $\F$ is said to be $(\rho,L)$-average-radius list decodable  if for every $y \in \F^n$ there are no $L+1$ codewords $c_1,c_2,\hdots,c_{L+1} \in \mathcal{C}$ such that
    $$\sum_{m \in [1:L+1]} wt(y - c_m) \leq (L+1)(\rho n),$$
    where $wt(.)$ denotes the Hamming weight.
\end{definition}
In \cite{VN14}, it was shown that if the code $\cal C$ is $(\rho,L)$-average-radius list decodable, then it is $(\rho,L)$-list decodable.

Codes that achieve this generalized Singleton bound for average-radius list decoding are identified as higher order generalizations of MDS codes, in the recent work \cite{Rot22}.
\begin{definition}(List decodable MDS codes \cite{Rot22})
    Let $\mathcal{C}$ be an $(n,k)$-code over a field $\F$ with $q$ elements. For $L<q$, we say that $\mathcal{C}$ is list decodable-MDS$(L)$ (in short LD-MDS$(L)$), if $\mathcal{C}$ is $(\rho, L)$-average-radius list decodable for $$\rho = \dfrac{L}{L+1}\left(1-\dfrac{k}{n}\right).$$
\end{definition}

In other words, for any $y \in \F^n$, if $\cal C$ is a LD-MDS($L$) code, there do not exist $L+1$ distinct codewords $c_0,c_1,\hdots,c_L \in \mathcal{C}$ such that
$$\sum_{i=0}^{L} wt(c_i -  y) \leq (L+1) \rho n = L(n-k),$$
 where $wt(.)$ denotes the Hamming weight.

Note that the usual MDS codes are LD-MDS($1$).
In \cite{BGM22a}, a different notion of higher order MDS codes was introduced as a generalization of MDS codes, using the notion of generic matrices. Such codes are potentially useful in designing codes for distributed storage \cite{BGM22}. Formally, these codes are referred to as \textit{higher order MDS} codes. The definition of such codes is as follows. 
\begin{definition}{(Higher order MDS codes \cite{BGM22a})}
    \label{defn:higherordermds}
   For a positive integer $\ell$, we say that $\mathcal{C}$ is $(n,k)$-MDS$(\ell)$ if it has dimension $k$, length $n$, and a generator matrix $G$ such that for any $\ell$ subsets $A_1,\hdots,A_{\ell} \subseteq [n]$ of size of at most $k$, we have that 
    $$dim(G_{A_1}\cap\cdots\cap G_{A_{\ell}}) = dim(W_{A_1}\cap \cdots \cap W_{A_\ell}),$$
    where $W_{k \times n}$ is a generic matrix over the same field characteristic and $G_{A_i}$, $W_{A_i}$ denote the span of columns of $G$ and $W$ indexed by $A_i$ respectively.
\end{definition}

Observe that the usual MDS codes are MDS$(\ell)$ for $\ell=1,2$.
In \cite{BGM22a}, it was shown that a code is MDS($\ell$) if and only if its dual is LD-MDS($L$) for all $L\leq \ell-1$. In a remarkable development, the work \cite{BGM22a} showed that generic RS codes are optimally list decodable, using this dual relationship. This implies that there exist RS codes (with evaluation points in a large enough field) that achieve the list-decoding capacity, i.e., meet the Generalized Singleton bound for list-decoding with equality. Many questions remain open however, regarding the minimum field size required for the existence of higher order MDS codes. The best known lower bound on the field size of an $(n,k)$-MDS($\ell$) code is $\Omega_{\ell,k}(n^{min\{\ell,k,n-k\}-1})$ \cite{BGM22}, whereas the best known (non-explicit) upper bound is $n^{O(min\{k,n-k\}(\ell-1))}$ \cite{BGM22,BGM22a,KMG21} which is exponential in the dimension. In \cite{Sto20,Rot22}, explicit constructions for MDS($3$) codes are presented over fields of sizes $2^{k^n}$ and $n^{k^{O(k)}}$ respectively, which are double exponential in $n$ for $k=\Theta(n)$. In a recent work\cite{BDG22}, the GM-MDS theorem \cite{Lov18} is used to give an explicit construction for $(n,k)$-MDS($\ell$) codes over fields of size $n^{(\ell k)^{O(\ell k)}}$. Further, explicit constructions for $(n,k)$-MDS(3) for $k=3,4,5$ over field sizes $O(n^3)$, $O(n^7)$, $O(n^{50})$ respectively are also provided in \cite{BDG22}.

\subsection{Our Contributions}

In this work, we show some new structural results relating to higher order MDS codes. The contributions and organization of this paper are as follows.
\begin{itemize}
    \item Using some combinatorial arguments, we show that an $(n,k)$-MDS($k$) code is also $(n,k)$-MDS($\ell$) for all $\ell >k$ (Please see Section \ref{sec:structResultsonHigherOrderMDS}, Lemma \ref{theorem:nkMDSktoMDSl}). This essentially reduces the problem of designing $(n,k)$-MDS($\ell$) codes to only those scenarios when $\ell\leq k$. This also enables us to make the bounds known for MDS($\ell$) independent of $\ell$, whenever $\ell\geq k$. 
    \item We also simplify the conditions required to be satisfied by an $(n,k)$-MDS($k-1$) code, for it to be an MDS($k$) code also (Please see Section \ref{sec:structResultsonHigherOrderMDS}, Lemma \ref{lemma:k-1genericenoughforMDS(k)}). This potentially reduces the complexity of constructing new MDS($k$) codes, from existing higher order MDS codes. 
    \item For certain parameter regimes, we show that $(n,k)$-MDS($k$) codes which are also RS codes (generated by a Vandermonde matrix defined on a set of evaluation points) are closed under the expurgation operation. In other words, the removal of the last row of the generator matrix of such codes results in an $(n,k-1)$-MDS($k-1$) code, for certain parameter regimes (Please see Section \ref{sec:vandBasedCodes}, Lemma \ref{lemma:removingrow}). We extend this to MDS($\ell$) codes for general $\ell$, under some special conditions (Please see Section \ref{sec:vandBasedCodes}, Lemma \ref{lemma:removingrowvand}).
    \item For $(n,k)$-MDS($k$) RS codes, we show the preservation of the higher order MDS property on a \textit{pseudo-shortening} operation (combining puncturing and expurgation). Specifically, given the Vandermonde generator matrix of an $(n,k)$-MDS($k$) RS code, we show that removing the last row and any column results in $(n-1,k-1)$-MDS($k-1$) code (Please see Section \ref{sec:vandBasedCodes}, Lemma \ref{lemma:sortofshortening}).
    \item Using well-known probabilistic tools, we obtain a new field size upper bound for the existence of $(n,k)$-MDS($\ell$) codes which is arguably better than existing bounds for certain parameter regimes (Please see Section \ref{sec:fieldSizeBounds}, Theorem \ref{theorem:fieldsizeboundnew}).
\end{itemize}



\section{Previous Work}
\label{sec:previous work}
To set the stage for our work, we recall a brief collection of existing results for higher order MDS codes which will be further used in deriving our results.

Towards recalling the previous results, the following definition of generic collections of sets is useful.
\begin{definition}
\label{defn:nklgenericcollection}
Let $\ell$ be a positive integer and let $n,k$ be integers such that $n\geq k \geq 0$. Consider $A_1,\hdots,A_{\ell} \subseteq [n]$ with $|A_i| \leq  k$. These sets  $(A_1,\hdots,A_{\ell})$ are called an $(n,k,\ell)$-generic collection if for all partitions $P_1 \cup \cdots\cup P_s = [ \ell ]$ we have
  \begin{equation}
  \label{eq:nklgenericcollection}
        \sum_{i=1}^{s}\left |\bigcap_{j \in P_i} A_j \right | \leq (s-1)k.
    \end{equation}
\end{definition}

One of the basic properties of higher order MDS codes is stated in \cite{BGM22} as:
\begin{lemma}
\label{lemma:basicpropoerty}
    \cite{BGM22} Let $\mathcal{C}$ be an $(n,k)$-MDS$(\ell)$ code. If $\ell \geq 3$, then $\mathcal{C}$ is also an MDS($\ell - 1$) code.
\end{lemma}

Observe that the property of a code being MDS($\ell$) is not dependent on any particular generator matrix. That is, if there is a generator matrix of the code satisfying the conditions of Definition \ref{defn:higherordermds}, then the conditions are satisfied by every generator matrix of the same code. Hence, throughout the paper, we alternate between saying that a matrix possesses the MDS($\ell$) property, or that the code (generated by the matrix) has the MDS($\ell$) property. 

The condition for a $k \times n$ matrix to be MDS$(\ell)$ in terms of the intersection of submatrices is given in \cite{BGM22} as:
\begin{lemma}
\label{lemma:nkmdsl}
    \cite{BGM22} Let $V$ be a $k \times n$ matrix. Then $V$ is $(n,k)$-MDS($\ell$) if and only if for all $(n,k,\ell)$-generic collections $(A_1,\hdots,A_{\ell})$ with $|A_1|+\cdots+|A_{\ell}| = ({\ell}-1)k$, we have that $V_{A_1} \cap \cdots \cap V_{A_{\ell}} = \B 0.$
\end{lemma}
An equivalent determinant-based criterion was also derived, which we recall.
\begin{lemma}
\label{lemma:vtodetcondition}
\cite{Tia19,BGM22} Let $V$ be a $k\times n$ matrix. Consider $A_1,A_2,\hdots,A_\ell\subseteq [n]$ with $|A_i|\leq k$ and $|A_1|+\cdots+|A_\ell|=(\ell-1)k$, we have that $V_{A_1}\cap \cdots \cap V_{A_\ell}= \B 0$ if and only if
\begin{equation}
\label{eq:vtodetcondition}
\det\begin{pmatrix}
I_k & V_{A_1} & & &\\
I_k & & V_{A_2} & &\\
\vdots & & & \ddots &\\
I_k & & & & V_{A_{\ell}}
\end{pmatrix}\neq 0,
\end{equation}
where $V_{A_i}$ denotes the submatrix of $V$ with columns indexed by $A_i$.
\end{lemma}
For some $\beta_i\in\F,\forall i\in[n]$, denote the $k\times n$ Vandermonde matrix $$\vand_{k}(\{\beta_i:i\in[n]\})\triangleq \begin{pmatrix}1&1&\cdots&1\\
\beta_1&\beta_2&\cdots&\beta_n\\
\vdots&\vdots&\cdots&\vdots\\
\beta_1^{k-1}&\beta_2^{k-1}&\cdots&\beta_n^{k-1}
\end{pmatrix}.$$
For the specific case of $(n,3)$-RS codes, we have the following simplified determinant conditions from \cite{BGM22,Rot22}.
\begin{lemma}
\label{lemma:n3MDS3}
    \cite{BGM22,Rot22} Let $V$ be $(n,3)$-RS code generated using $G=\vand_{3}(\{\beta_i:i\in[n]\})$. Then $V$ is MDS$(3)$ if and only if for all injective maps $\alpha : [6]\rightarrow[n]$ we have that
    $$\det\begin{pmatrix}
     1 & \beta_{\alpha(1)}+\beta_{\alpha(2)} & \beta_{\alpha(1)}\beta_{\alpha(2)}\\ 
     1 & \beta_{\alpha(3)}+\beta_{\alpha(4)} & \beta_{\alpha(3)}\beta_{\alpha(4)}\\
     1 & \beta_{\alpha(5)}+\beta_{\alpha(6)} &  \beta_{\alpha(5)}\beta_{\alpha(6)}
\end{pmatrix}
    \neq 0.$$
\end{lemma}
In \cite{BDG22}, an alternative characterization of the higher-order MDS conditions for Reed-Solomon codes is presented.
For $A\subseteq [n]$, define the polynomial $$\pi_{A}(x) \triangleq \prod_{i\in A}(x-\beta_i).$$
Define $\pi_{A}^{d}(x)$ to be the following (row) vector of polynomials:
$$\pi_{A}^{d}(x)\triangleq(\pi_A(x), x\pi_A(x), \hdots, x^{d-1}\pi_A(x)).$$ 
Then we have the following determinant condition from \cite{BDG22}.
\begin{lemma}
\label{lemma:detconditionmdsl}
    \cite{BDG22} Assume that $A_1,..A_\ell\in [n]$ such that $|A_i|\leq k, \forall i\in [\ell]$. Let $|A_1|+\cdots+|A_\ell|=(\ell-1)k$. Let $\delta_{i}=k-|A_i|$. Assume WLOG that $A_1=\{1,2,\hdots,k-\delta_1\}$. We have that $V_{A_1}\cap \cdots \cap V_{A_\ell}= \B 0$ iff 
    $$\det\begin{pmatrix}
    \pi_{A_{2}}^{\delta_2}(\beta_1) & \pi_{A_{3}}^{\delta_3}(\beta_1) & \cdots & \pi_{A_{\ell}}^{\delta_{\ell}}(\beta_1)\\
    \pi_{A_{2}}^{\delta_2}(\beta_2) & \pi_{A_{3}}^{\delta_3}(\beta_2) & \cdots & \pi_{A_{\ell}}^{\delta_\ell}(\beta_2)\\
    \vdots & \vdots & \cdots & \vdots \\
    \pi_{A_{2}}^{\delta_2}(\beta_{k-\delta_1}) & \pi_{A_{2}}^{\delta_3}(\beta_{k-\delta_1}) & \cdots & \pi_{A_{\ell}}^{\delta_\ell}(\beta_{k-\delta_1})
    \end{pmatrix}\neq 0.$$
\end{lemma}
We recall an equivalence between LD-MDS codes and the dual of higher order MDS codes from \cite{BGM22a}, which implies that our results are effectively applicable for LD-MDS codes also. 
\begin{proposition}\cite{BGM22a}
If $\mathcal{C}$ is a linear code then for all $\ell \geq 1$, $\mathcal{C}$ is MDS($\ell + 1$) if and only if $\mathcal{C}^{\perp}$ is LD-MDS($\leq \ell$).
\end{proposition}

\section{On the Structure of Higher Order MDS Codes}
\label{sec:structResultsonHigherOrderMDS}
In this section, we present some structural results for higher order MDS codes. We show that an $(n,k)$-MDS($k$) code is also an $(n,k)$-MDS($\ell$) code for all $\ell >k$ using some combinatorial arguments. We also obtain simplified conditions (as compared to Lemma \ref{lemma:nkmdsl}) required to be satisfied by an $(n,k)$-MDS($k-1$) code, for it to be an $(n,k)$-MDS($k$) code also. We now give our first main result. 
\begin{theorem}
\label{theorem:nkMDSktoMDSl}
    If $V$ is an $(n,k)$-MDS($k$) code, then $V$ is also $(n,k)$-MDS($\ell$) code $\forall\ \ell \geq k$.
\end{theorem}
\begin{IEEEproof}
    From Lemma \ref{lemma:nkmdsl}, we know that $V$ is $(n,k)$-MDS($\ell$) if and only if for all $(n,k,\ell)$-generic collection of subsets $(A_1,\hdots,A_\ell): A_i\subseteq[n],  \forall i$ with 
    \begin{equation}
    \label{eq:Aisuml-1k}
        |A_1|+\cdots+|A_{\ell}| = ({\ell}-1)k,
    \end{equation}
    we have that 
    \begin{equation}
    \label{eq:genericintersectionl}
        V_{A_1} \cap \cdots \cap V_{A_{\ell}} = \B 0.
    \end{equation}
    We prove that V is $(n, k)$-MDS($\ell$) for all $\ell \geq k$ by showing that (\ref{eq:genericintersectionl}) is satisfied for each possible $(n, k,\ell)$-generic collection satisfying (\ref{eq:Aisuml-1k}).
    
    Observe that when any of the $A_i$s are empty, then (\ref{eq:genericintersectionl}) holds trivially. Now, we look at all non-zero possibilities for the cardinality of all $A_i$s in an $(n,k,\ell)$-generic collection and check if (\ref{eq:Aisuml-1k}) holds:
    \begin{itemize}
        \item  Suppose all $A_i$s are of size $k$, then 
        $$|A_1|+\cdots+|A_{\ell}| = \ell k.$$
        But from Lemma \ref{lemma:nkmdsl}, we want $|A_1|+\cdots+|A_{\ell}| = ({\ell}-1)k$. So, all $A_i$s cannot be of size $k$.
        \item Suppose, exactly $(\ell - 1)$ of the $A_i$s are of size $k$. WLOG, say $A_1,...A_{\ell -1}$ are of size $k$. Then
        \begin{equation}
        \label{eq:l-1Aisk}
            |A_1|+\cdots+|A_{\ell - 1}| = (\ell - 1) k.
        \end{equation}
        From (\ref{eq:Aisuml-1k}) and (\ref{eq:l-1Aisk}), we get $|A_{\ell}| = 0$, and this scenario is already handled. So, we cannot have $(\ell - 1)$ of the $A_i$s of size $k$.
        \item For $2\leq x \leq k$, suppose $(\ell - x)$ of the $A_i$s are of size $k$. WLOG, say $A_1,\hdots, A_{\ell -x}$ are of size $k$. Then
        \begin{equation}
        \label{eq:l-xAisk}
            |A_1|+\cdots+|A_{\ell - x}| = (\ell - x) k.
        \end{equation}  
        From (\ref{eq:Aisuml-1k}) and (\ref{eq:l-1Aisk}), we get
        \begin{equation}
             |A_{\ell - x+1}| + \cdots + |A_{\ell}| = (x-1)k.
        \end{equation}
        Let $\{Q_1,\hdots,Q_{s'}\}$ be any arbitrary partition of $[\ell]\setminus[\ell-x]$. Consider the following partition of $[\ell]$ given as $\{P_i=\{i\}:i \in[\ell-x]\}\cup\{Q_1,\hdots,Q_{s'}\}$. As $\{A_1,\ldots,A_{\ell}\}$ is $(n,k,\ell)$-generic collection of subsets of $[n]$, we have that
        \begin{align}
        \nonumber
        \sum_{i=1}^{\ell - x}\left |\bigcap_{j \in P_i} A_j \right |+ \sum_{i=1}^{s'}\left |\bigcap_{j \in Q_i} A_j \right | &\leq (\ell - x+s'-1)k \\
        \nonumber
        (\ell - x)k + \sum_{i=1}^{s'}\left |\bigcap_{j \in Q_i} A_j \right | &\leq (\ell - x + s' - 1)k \\
        \label{eq:genericsetsx}
        \sum_{i=1}^{s'}\left |\bigcap_{j \in Q_i} A_j \right | &\leq (s'-1)k.
        \end{align}
        Using (\ref{eq:genericsetsx}) we can say that $A_{\ell-x+1},\hdots,A_{\ell}$ are a $(n,k,x)$-generic collection of subsets of $[n]$. As $V$ is MDS($k$) it is also MDS($x$) as $x\leq k$, which implies $V_{A_{\ell - x+1}} \cap \cdots \cap V_{A_{\ell}} = \B 0$. Therefore $V_{A_1} \cap \cdots \cap V_{A_{\ell}} = \B 0$.
        \item For $k<x\leq \ell$, if $(\ell-x)$ of the $A_i$s are of size $k$, WLOG say $A_1,\hdots,A_{\ell -x}$ are of size $k$. Then 
        \begin{equation}
        \label{eq:Aisuml-xk}
             |A_1|+\cdots+|A_{\ell - x}| = (\ell - x) k.
        \end{equation}  
        Using (\ref{eq:Aisuml-1k}) and (\ref{eq:Aisuml-xk}), we get
            $$|A_{\ell - x+1}| + \cdots + |A_{\ell}| = (x-1)k = xk-k.$$
         But,
            $$|A_{\ell - x+1}| + \cdots + |A_{\ell}| \leq x(k-1) = xk-x$$
        (as each $A_i$, $i\in [\ell]\setminus [\ell-x]$ is of size at most $k-1$). As $xk-x < xk-k$, we get a contradiction. Hence, we cannot have $(\ell-x)$ $A_i$s of size $k$ when $x>k$.
    \end{itemize}
   From above, we have that for all $A_1,\hdots,A_{\ell} \subseteq [n]$ with $|A_i| \leq  k$  and $|A_1|+\cdots+|A_{\ell}| = ({\ell}-1)k$, we have $V_{A_1} \cap \cdots \cap V_{A_{\ell}} = \B 0$ as long as $V$ is $(n,k)$-MDS($k$). Hence, $V$ is $(n,k)$-MDS($\ell$) code $\forall\ \ell \geq k$.
\end{IEEEproof}
From Lemma \ref{lemma:basicpropoerty}, we have that, an $(n,k)$-MDS$(\ell)$ code $\mathcal{C}$ is also $(n,k)$-MDS$(\leq \ell)$, $\forall\ \ell \geq 3$. Therefore, from Lemma \ref{lemma:basicpropoerty} and Theorem \ref{theorem:nkMDSktoMDSl} we have the following corollary.
\begin{corollary}
\label{corollary:nkMDSktomdsliff}
For $\ell \geq k$, an $(n,k)$ code $\mathcal{C}$ is  $(n,k)$-MDS$(\leq \ell)$
if and only if it is $(n,k)$-MDS($k$).
\end{corollary}
 From Lemma \ref{lemma:n3MDS3} and Theorem \ref{theorem:nkMDSktoMDSl}, we get the following simple observation for $k=3$.
\begin{observation}
   Let $V$ be an $(n,3)$-RS code with evaluation points $\beta_1,\ldots,\beta_n \in \mathbb{F}$. Then $V$ is $(n,3)$-MDS$(\ell)$, $\ell\geq 3$ if and only if for all injective maps $\alpha : [6]\rightarrow[n]$ we have that
    $$\det\begin{pmatrix}
     1 & \beta_{\alpha(1)}+\beta_{\alpha(2)} & \beta_{\alpha(1)}\beta_{\alpha(2)}\\ 
     1 & \beta_{\alpha(3)}+\beta_{\alpha(4)} & \beta_{\alpha(3)}\beta_{\alpha(4)}\\
     1 & \beta_{\alpha(5)}+\beta_{\alpha(6)} &  \beta_{\alpha(5)}\beta_{\alpha(6)}
\end{pmatrix}
    \neq 0.$$
\end{observation}

Towards obtaining the second result in this section, we provide the following definition for $(n,k,\ell)_{k-1}$-generic collections.
\begin{definition}
    A $(n,k,\ell)$-generic collection $(A_1,\hdots,A_\ell)$ is said to be a $(n,k,\ell)_{k-1}$-generic collection if $|A_i|=k-1, \forall i$.
\end{definition}

Note that the subsets in a $(n,k,k)_{k-1}$-generic collection must be distinct as the intersection of any two subsets in this collection have at most $(k-2)$ elements in common (a property that can be verified using Definition \ref{defn:nklgenericcollection}). 

Now, we show that for an $(n,k)$-MDS$(k-1)$ code to be an MDS($k$) code, it suffices to look only at all possible $(n,k,k)_{k-1}$-generic collections, which is simpler as compared to the conditions of Lemma \ref{lemma:nkmdsl}. 
\begin{lemma}
\label{lemma:k-1genericenoughforMDS(k)}
Let $V$ be an $(n,k)$-MDS$(k-1)$ code. Then $V$ is MDS($k$) if and only if for each $(n,k,k)_{k-1}$-generic collection $(A_1,\hdots,A_k)$, we have $\cap_{i=1}^{k}V_{A_i}=\B{0}$. 
\end{lemma}
The proof of Lemma \ref{lemma:k-1genericenoughforMDS(k)} is provided in Appendix \ref{proof:k-1genericenoughforMDS(k)}.
%

\section{On Higher Order MDS Codes Based on RS Codes}
\label{sec:vandBasedCodes}
In this section we present the results that concern the properties of higher order MDS codes based on Reed-Solomon codes generated from Vandermonde matrix. We show that $(n,k)$-MDS($k)$ RS codes are closed under expurgation operation under certain parameter regimes, and extend this to MDS($\ell$) codes for general $\ell$. We also show the preservation of the higher order MDS property on a pseudo-shortening operation on the generator matrix of a RS $(n,k)$-MDS($k$) code. 

In the following Lemma, we show that removing the last row (expurgation operation) from the generator matrix of an MDS$(k)$ Reed-Solomon code results in an MDS$(k-1)$ code.

\begin{lemma}
\label{lemma:removingrow}
Let $V$ be an $(n,k)$-RS code generated using $G=\vand_{k}(\{\beta_i:i\in[n]\})$ such that $V$ is $(n,k)$-MDS($k$) and $n\geq (k-2)(k-1)+1$. Then the code $V'$ generated by $\vand_{k-1}(\{\beta_i:i\in[n]\})$ is an $(n,k-1)$-MDS($k-1$) code. 
\end{lemma}
The proof of Lemma \ref{lemma:removingrow} is provided in Appendix \ref{proof:removingrow}.

Lemma \ref{lemma:removingrowvand} extends the result of Lemma  \ref{lemma:removingrow}  to more general $\ell$, under some conditions.
\begin{lemma}
\label{lemma:removingrowvand}
    Let $V$ be $(n,k)$-RS code generated using $G=\vand_{k+1}(\{\beta_i:i\in[n]\})$ such that $V$ is $(n,k+1)$-MDS($\ell$) and $n\geq (\ell-1)k+1$. Then the code $V'$ generated by $\vand_{k}(\{\beta_i:i\in[n]\})$ is an $(n,k)$-MDS($\ell$) code.
\end{lemma}
The proof of Lemma \ref{lemma:removingrowvand} is provided in Appendix \ref{proof:removingrowvand}.

In the following lemma we show that removing the last row and any column  (expurgation and puncturing together, an operation we have called pseudo-shortening) from the generator matrix of a MDS$(k)$ RS code also results in an MDS$(k-1)$ code. 
\begin{lemma}
\label{lemma:sortofshortening}
Let $V$ be $(n,k)$-RS code generated using $G=\vand_{k}(\{\beta_i:i\in[n]\})$ such that $V$ is $(n,k)$-MDS($k$). Then, for any $j\in[n]$, the code $V'$ generated by $\vand_{k-1}(\{\beta_i:i\in[n]\}\setminus \beta_j)$ is an $(n-1,k-1)$-MDS($k-1$) code. 
\end{lemma}
The proof of Lemma \ref{lemma:sortofshortening} is provided in the Appendix \ref{proof:sortofshortening}.
\section{A New Field Size Bound}
\label{sec:fieldSizeBounds}
In this section, we show a new upper bound for the field size of $(n,k)$-MDS$(\ell)$ codes. To do this, we essentially rely upon some probabilistic tools, specifically the Lovasz's Local Lemma and Schwarz Zippel Lemma, which we now recall.


\begin{lemma}{\textbf{(Schwarz Zippel Lemma)}}
\label{lemma:SZ}
    Let $p(x_1, \hdots, x_n)$ be a nonzero polynomial of $n$ variables with degree $d$ with coefficients from a field $\F$. Let $S$ be a finite subset of $\F$, with at least $d$ elements in it. If we assign $x_1,\hdots,x_n$ values from $S$ independently and uniformly at random, then
    $$Pr[p(x_1,\hdots,x_n) = 0] \leq \dfrac{d}{|S|}.$$
\end{lemma}
\begin{lemma}{(\textbf{Lovasz Local Lemma} \cite{lovaszlocallemma})}
\label{lemma:locallemma}
Let $B_1,B_2,\hdots,B_n$ be events in an arbitrary probability space. Suppose that each event $B_i$ is mutually independent of a set of all other events $B_j$ but at most $d$, and that $Pr(B_i) \leq p$ for all $1 \leq\ i \leq n.$ If $$ep(d+1) \leq 1$$ then $Pr(\bigcap_{i=1}^{n} \bar{B_i}) > 0$, where $\bar{B_i}$ denotes the complement of event $B_i$, and $e$ denotes the base of the natural logarithm. 
\end{lemma}
In the context of Lemma \ref{lemma:locallemma}, we use the phrase `bad' events to represent the events $B_i$. Thus, what Lemma \ref{lemma:locallemma} assures us is that if the probability of any bad event is small enough and any bad event is not dependent on too many others, then there is a positive probability that none of them occur.

We are now ready to present our results. Let $\mathfrak A$ denote the set of all collection of sets $\underline{A}=\left\{\{A_1,A_2,\hdots,A_{\ell}\}: A_j\subset[n], \forall j\right\}$ that satisfy the condition in Lemma \ref{lemma:vtodetcondition}. Throughout this section, we consider that the matrix $V=\vand_{k}(\{\beta_i:i\in[n]\})$.
\begin{claim}
\label{claim:vdetdegree}
The determinant in (\ref{eq:vtodetcondition}) in Lemma \ref{lemma:vtodetcondition} corresponding to the matrix $V$, as a polynomial in $\beta_i:i\in[n]$, has total degree $\leq \ell k^2$. 
\end{claim}
The proof of Claim \ref{claim:vdetdegree} is given in Appendix \ref{proof:vdetdegree}.



For $\underline{A}=\{A_1,\hdots,A_\ell\}, \underline{A}'=\{A_1',\hdots,A_\ell'\}\in \mathfrak{A}$, we say that $\underline{A}    \not\perp\!\!\!\perp
\underline{A}'$ if there is at least one $s\in\left(\cup_j^\ell A_j\right)\cap (\cup_j^\ell A_j')$. We then say that $\underline{A}$ and $\underline{A}'$ are dependent (on each other). For $\underline{A}\in\mathfrak{A}$, let $E_{\underline{A}}$ denote the event that the determinant given by (\ref{eq:vtodetcondition}) in Lemma \ref{lemma:vtodetcondition} is zero, for the collection $\underline{A}$. We then have the following claim.
\begin{claim}
\label{claim:qcondforMDS}
If 
$q\geq e\cdot\ell k^2\cdot\left(\max_{\underline{A}\in\mathfrak{A}}|\{\underline{A}'\in\mathfrak{A}\colon \underline{A}\not\perp\!\!\!\perp\underline{A}'\}|\right)$, then there exist evaluation points $\beta_i:i\in[n]$ such that the code generated by $\vand_{k}(\{\beta_i:i\in[n]\})$  is  MDS($\ell$).
\end{claim}
\begin{IEEEproof}
We use the Lemma \ref{lemma:locallemma} with the bad events being the events $E_{\underline{A}}:\underline{A}\in\mathfrak{A}$. By Claim \ref{claim:vdetdegree} and the Schwarz Zippel Lemma, the probability that the determinant (\ref{eq:vtodetcondition}) in Lemma \ref{lemma:vtodetcondition} is $0$, for any given $\underline{A}\in\mathfrak{A}$, when choosing the evaluation points independently and uniformly at random from field $\F$ (with $q$ elements) is at the most $\frac{\ell k^2}{q}$. Thus, the probability of any bad event $E_{\underline{A}}$ is at the most $\frac{\ell k^2}{q}$. Invoking Lemma \ref{lemma:locallemma}, the proof is complete. 
\end{IEEEproof}
The following claim gives an upper bound to the quantity $\max_{\underline{A}\in\mathfrak{A}}|\{\underline{A}'\in\mathfrak{A}\colon \underline{A}\not\perp\!\!\!\perp\underline{A}'\}|$ in Claim \ref{claim:qcondforMDS}. 
\begin{claim}
\label{claim:noofAcond}
For any $\underline{A}\in{\mathfrak A}$, we have
\begin{equation*}
|\{\underline{A}'\in\mathfrak{A}\colon \underline{A}\not\perp\!\!\!\perp\underline{A}'\}|\leq \min\left(2^n,\Delta \left(\frac{en}{\Delta/\ell}\right)^{\Delta}\right) \cdot \min\left(2^{\ell\min(\Delta,n)},k^\ell\min(\Delta,n)^{k\ell}\right).
\end{equation*}
\end{claim}
The proof of Claim \ref{claim:noofAcond} is given in Appendix \ref{proof:noofAcond}.
\begin{theorem}
\label{theorem:fieldsizeboundnew}
There exists an $(n,k)$-MDS$(\ell)$ code over a field of size $q$, if $$\scriptsize q\geq e\ell k^2\cdot\min\left(2^n,\Delta \left(\frac{en}{\Delta/\ell}\right)^\Delta\right)\cdot\min\left(2^{\ell\min(\Delta,n)},k^\ell\min(\Delta,n)^{k\ell}\right).$$
\end{theorem}
\begin{IEEEproof}
\label{proof:fieldSizeBoundproof}
    From Claim \ref{claim:qcondforMDS} we have that, if
    $$q\geq e.\ell k^2.\left(\max_{\underline{A}\in\mathfrak{A}}|\{\underline{A}'\in\mathfrak{A}\colon \underline{A}\not\perp\!\!\!\perp\underline{A}'\}|\right)$$ then there exists a choice of evaluation points $\beta_i\in \F:i\in[n]$ ($\F$ being a field of size $q$) such that $V$ is $(n,k)$-MDS$(\ell)$. Using Claim \ref{claim:qcondforMDS}, Theorem \ref{theorem:fieldsizeboundnew} follows. 
    
\end{IEEEproof}
\begin{remark}
A previously known upper bound from \cite{Rot22,BGM22a,BGM22} was
\begin{equation*}
    q \geq \ell n^2\binom{n}{\leq k}^\ell= n^{{\cal O}(\ell\min(k,n-k))}.
\end{equation*}
However, note that $\binom{n}{\leq k}\leq 2^n$. Hence, we see that this existing bound is essentially  
\begin{equation*}
    \min(2^{{\cal O}(\ell n))},n^{{\cal O}(\ell\min(k,n-k))}).
\end{equation*}
Observe that our new bound in Theorem \ref{theorem:fieldsizeboundnew} arguably tightens this bound in some special regimes, when $\Delta=(\ell-1)k<n$. Indeed, when $\Delta<n$, our bound is at the most ${\cal O}\left(\ell^2 k^3\left(\frac{n}{k}\right)^{\ell k}2^{\ell^2k}\right)={\cal O}\left(\ell^2 k^3\left(\frac{2^\ell n}{k}\right)^{\ell k}\right)$, which is tighter than ${\cal O}(n^{\ell k})$ for small values of $\ell$ (as compared to $k$).
\end{remark}

%

\section{Conclusion}
\label{sec:conclusion}
In this work, we have derived some structural results for higher order MDS codes. We showed that higher order MDS codes of dimension and order $k$ are also higher order MDS codes of order $\ell$ for any $\ell>k$.  We showed performing expurgating and shortening-like operations on higher order MDS codes obtained from Reed-Solomon codes (for certain parameter regimes) result in new higher order MDS codes with a related set of parameters. We also showed, using the Local lemma, the existence of higher order MDS codes  requiring smaller field size (for certain parameter regimes) than the best known upper bound. We believe that these results will aid in constructing efficient higher order MDS codes, perhaps by bootstrapping on existing codes. 


\bibliographystyle{IEEEtran} 
\bibliography{Bibilography} 








\clearpage
\begin{appendices}

\section{Proof of Lemma \ref{lemma:k-1genericenoughforMDS(k)}}
\begin{IEEEproof}
\label{proof:k-1genericenoughforMDS(k)}
   From Lemma \ref{lemma:nkmdsl}, we know that $V$ is $(n,k)$-MDS($k$) if and only if for all $(n,k,k)$-generic collection of subsets $A_1,\ldots,A_{k}$ with
   \begin{equation}
   \label{eq:kgenericsum}
       |A_1|+\cdots+|A_{k}| = (k-1)k,
   \end{equation}
   we have that
   \begin{equation}
   \label{eq:kgenericintersection}
   V_{A_1} \cap ... \cap V_{A_{k}} = \B 0.
   \end{equation}
Now, we want to show that, given $V$ is $(n,k)$-MDS($k-1$), it suffices to check $V_{A_1} \cap ... \cap V_{A_{k}} = \B 0$ when $A_i$s are a $(n,k,k)_{k-1}$-generic collection for $V$ to be $(n,k)$-MDS($k$).
   
    Observe that when any of the $A_i\colon i \in [k]$ are empty, then (\ref{eq:kgenericintersection}) holds trivially. Now, if we look at all non-zero possibilities for the cardinality of $A_i$s in a generic collection and check if (\ref{eq:kgenericintersection}) holds:
    \begin{itemize}
        \item  Suppose all $A_i$s are of size $k$, then 
        $$|A_1|+\cdots+|A_{\ell}| = \ell k.$$
        But from Lemma \ref{lemma:nkmdsl}, we want $|A_1|+\cdots+|A_{\ell}| = ({\ell}-1)k$. So, all $A_i$s cannot be of size $k$.
        \item Suppose, exactly $(k - 1)$ of the $A_i$s are of size $k$, WLOG say $A_1,...A_{k -1}$ are of size $k$. Then
        \begin{equation}
            |A_1|+\cdots+|A_{k - 1}| = (k - 1) k.
        \end{equation}
        From (\ref{eq:kgenericsum}) and (\ref{eq:kgenericintersection}), we get $|A_{k}| = 0$, a contradiction to our assumption that none of the $A_i:i\in[k]$ have size $0$. Thus, we cannot have $(k - 1)$ of the $A_i$s of size $k$.
        \item For $2\leq x \leq k-1$, suppose $(k - x)$ of the $A_i$s are of size $k$, WLOG say $A_1,...A_{k -x}$ are of size $k$. Then 
        \begin{equation}
        \label{eq:kxgenericsum}
            |A_1|+\cdots+|A_{k - x}| = (k - x) k
        \end{equation}  
        From (\ref{eq:kgenericsum}) and (\ref{eq:kxgenericsum}), we get
        \begin{equation}
             |A_{k - x+1}| + .. + |A_k| = (x-1)k.
        \end{equation}
        Let $\{Q_1,\hdots,Q_{s'}\}$ be any arbitrary partition of $[k]\setminus[k-x]$. Consider the following partition of $[k]$ given as $\{P_i=\{i\}:i \in[k-x]\}\cup\{Q_1,\hdots,Q_{s'}\}$. As $(A_1,\cdots A_{k})$ is an $(n,k,k)$-generic collection of subsets of $[n]$, we have that
        \begin{align}
        \nonumber
        \sum_{i=1}^{k - x}\left |\bigcap_{j \in P_i} A_j \right |+ \sum_{i=1}^{s'}\left |\bigcap_{j \in Q_i} A_j \right | &\leq (k - x+s'-1)k \\
        \nonumber
        (k - x)k + \sum_{i=1}^{s'}\left |\bigcap_{j \in Q_i} A_j \right | &\leq (k - x + s' - 1)k \\
        \label{eq:genericcondsatisfies}
        \sum_{i=1}^{s'}\left |\bigcap_{j \in Q_i} A_j \right | &\leq (s'-1)k.
        \end{align}
        Using (\ref{eq:genericcondsatisfies}) we can say that $(A_{k-x+1},...,A_{k})$ is an $(n,k,x)$-generic collection of subsets of $[n]$. As $V$ is MDS($k-1$) it is also MDS($x$), which implies $V_{A_{k - x+1}} \cap .. \cap V_{A_{k}} = \B 0$. Therefore $V_{A_1} \cap ... \cap V_{A_{k}} = \B 0$ is satisfied automatically by the given condition.
         \item Suppose, there are no $A_i:i\in[k]$ of size $k$. Then, each of the $A_i:i\in[k]$ are of size $\leq (k -1)$. In order to satisfy (\ref{eq:kgenericsum}) the only possibility is each $A_i:i\in[k]$ must have cardinality equal to $k-1$. Note that this means $(A_i:i\in[k])$ is an $(n,k,k)_{k-1}$-generic collection.
         Therefore, for $V$ to be $(n,k)$-MDS$(k)$ it suffices to check $V_{A_{1}} \cap ... \cap V_{A_{k}} = \B 0$ for each $(n,k,k)_{k-1}$-generic collection $(A_1,\hdots,A_k)$.
    \end{itemize}
\end{IEEEproof}

\section{Proof of Lemma \ref{lemma:removingrow}}
\begin{IEEEproof}
\label{proof:removingrow}
    Suppose $V$ is an $(n,k)$-MDS$(k)$ code but $V'$ is not $(n,k-1)$-MDS($k-1$) code. Then by Lemma \ref{lemma:nkmdsl}, there exists some $(n,k-1,k-1)$-generic collection $A_{1}^{'},..,A_{k-1}^{'} \subseteq [n]$, with
    \begin{equation}
    \label{eq:removingroweq}
        |A_{1}^{'}|+|A_{2}^{'}|+..+|A_{k-1}^{'}| = (k-2)(k-1),
    \end{equation}    
    such that $V_{A_{1}^{'}}^{'} \cap ... \cap V_{A_{k-1}^{'}}^{'} \neq \B 0$.  
    Now, we construct a $(n,k,k)$-generic collection $A_{1},..,A_{k} \subseteq [n]$, using 
    $A_{1}^{'},..,A_{k-1}^{'}$ in the following way. Let $p \in [n]$ and $p \notin (A_1^{'}\cup .. \cup A_{k-1}^{'})$. Such a $p$ exists as $n\geq (k-2)(k-1)+1$. Consider, $A_{i}=A_{i}^{'}\cup \{p\}, \forall i\in [k-1]$ and let $A_{k}=A_{x}^{'}\cup$\{one element from $(A_{1}^{'}\cup ..\cup A_{k-1}^{'})\setminus A_{x}^{'}$\}, where $A_{x}^{'}$ is of size $(k-2)$ ( such an $A'_x$ of size $(k-2)$ exists, otherwise (\ref{eq:removingroweq}) will not be satisfied). Now, $|A_{i}| \leq k-1+1=k$, $i\in[k-1], |A_k|=k-1$. From definition (\ref{defn:nklgenericcollection}), for all partitions $P_1 \cup ...\cup P_s = [k-1]$, we have,
        $$\sum_{i=1}^{s}\left |\bigcap_{j \in P_i} A_{j}^{'} \right | \leq (s-1)(k-1).$$
    As $A_{i}=A_{i}^{'}\cup \{p\}, \forall i\in [k-1]$, we get
        \begin{align}
        \label{eqn1001}\sum_{i=1}^{s}\left |\bigcap_{j \in P_i} A_{j} \right | \leq (s-1)(k-1)+s = sk-k+1.
        \end{align}
    Now consider a partition $Q_i:i\in[s']$ of $[k]$. Suppose that there exists a $Q_{i'}$ such that element $k\in Q_{i'}$ and $|Q_{i'}|\geq 2$. Then, the collection $\{Q_i:i\in[s']\setminus\{i'\}\}\cup\{Q_{i'}\setminus \{k\}\}$ forms a partition of $[k-1]$. Since $A_k$ does not contain the element $p$, by (\ref{eqn1001}) we have that 
    \begin{equation}
    \label{eq:def11}
       \sum_{i=1}^{s'}\left |\bigcap_{j \in Q_i} A_{j} \right | \leq s'k-k =(s'-1)k.
    \end{equation}
    If there however is no such $Q_{i'}$ such that $k\in Q_{i'}$ and $|Q_{i'}|\geq 2$, then there should be a $Q_{i''}$ such that $k\in Q_{i''}$ and $|Q_{i''}|=1$. In that case, the collection $\{Q_i:i\in[s']\}\setminus \{Q_{i''}\}$ forms a partition of $[k-1]$. Hence, again using (\ref{eqn1001}), we see that 
    \begin{align}
    \nonumber
         \sum_{i=1}^{s'}\left |\bigcap_{j \in Q_i} A_{j} \right |&= \sum_{i=1\colon i\neq i''}^{s'}\left |\bigcap_{j \in Q_i} A_{j} \right | + \left|A_k\right|\\
         \label{eq:def12}
         &\leq (s'-1)k-k+1+(k-1)=(s'-1)k,
    \end{align}
    as $|A_k| = k-1$. 
    
    

    From equations (\ref{eq:def11}) and (\ref{eq:def12}) we can say that $A_1,A_2,..,A_k$ form a $(n,k,k)$-generic collection.

    We now show that for the above obtained $(n,k,k)$-generic collection, $\cap_{i=1}^k V_{A_i}\neq \B 0$, thus showing that the code $V$ is not MDS($k$), leading to a contradiction with the given statement. 
        
    From Lemma \ref{lemma:detconditionmdsl}, assuming WLOG $A_k=\{1,\hdots,k-1\}$, we have $V_{A_1} \cap ... \cap V_{A_k} = \B 0$ if and only if
    \begin{equation}
    \label{eq:detforgenericintersection}
    \det\begin{pmatrix}
    \pi_{A_{1}}^{\delta_1}(\beta_1) & \pi_{A_{2}}^{\delta_2}(\beta_1) & .. & \pi_{A_{k-1}}^{\delta_{k-1}}(\beta_1)\\
    \pi_{A_{1}}^{\delta_1}(\beta_2) & \pi_{A_{2}}^{\delta_2}(\beta_2) & .. & \pi_{A_{k-1}}^{\delta_{k-1}}(\beta_2)\\
    . & . & .. & . \\
    . & . & .. & . \\
    \pi_{A_{1}}^{\delta_1}(\beta_{k-1}) & \pi_{A_{2}}^{\delta_2}(\beta_{k-1}) & .. & \pi_{A_{k-1}}^{\delta_{k-1}}(\beta_{k-1})
    \end{pmatrix}\neq 0
    \end{equation}
    We now show that the determinant in (\ref{eq:detforgenericintersection}) is zero so that $V_{A_1} \cap ... \cap V_{A_k} \neq \B 0$ which implies that our assumption is incorrect.
    As $p\in (A_{1}\cap A_{2}\cap .. \cap A_{k-1})$, the  equation (\ref{eqn:5020}) holds, where $\delta_{i}^{'} = k-1-|A_{i}'|$.
    \begin{equation}
    \scriptsize
    \det\begin{pmatrix}
    \pi_{A_{1}}^{\delta_1}(\beta_1) & \pi_{A_{2}}^{\delta_2}(\beta_1) & .. & \pi_{A_{k-1}}^{\delta_{k-1}}(\beta_1)\\
    \pi_{A_{1}}^{\delta_1}(\beta_2) & \pi_{A_{2}}^{\delta_2}(\beta_2) & .. & \pi_{A_{k-1}}^{\delta_{k-1}}(\beta_2)\\
    . & . & .. & . \\
    . & . & .. & . \\
    \pi_{A_{1}}^{\delta_1}(\beta_{k-1}) & \pi_{A_{2}}^{\delta_2}(\beta_{k-1}) & .. & \pi_{A_{k-1}}^{\delta_{k-1}}(\beta_{k-1})
    \end{pmatrix}
    =  
    \prod_{j=1}^{k-1}(\beta_j - \beta_p)\times
    \det\begin{pmatrix}
    \pi_{A_{1}^{'}}^{\delta_{1}^{'}}(\beta_1) & \pi_{A_{2}^{'}}^{\delta_{2}^{'}}(\beta_1) & .. & \pi_{A_{k-1}^{'}}^{\delta_{k-1}^{'}}(\beta_1)\\
    \pi_{A_{1}^{'}}^{\delta_{1}^{'}}(\beta_2) & \pi_{A_{2}^{'}}^{\delta_{2}^{'}}(\beta_2) & .. & \pi_{A_{k-1}^{'}}^{\delta_{k-1}^{'}}(\beta_2)\\
    . & . & .. & . \\
    . & . & .. & . \\
    \pi_{A_{1}^{'}}^{\delta_{1}^{'}}(\beta_{k-1}) & \pi_{A_{2}^{'}}^{\delta_{2}^{'}}(\beta_{k-1}) & .. & \pi_{A_{k-1}^{'}}^{\delta_{k-1}^{'}}(\beta_{k-1})
    \end{pmatrix}\label{eqn:5020}
    \end{equation}
    Since $\beta_j\neq \beta_p, \forall j\in A_k$, the term $\prod_{j=1}^{k-1}(\beta_j - \beta_p)$ on RHS of (\ref{eqn:5020}), is non-zero. From Lemma \ref{lemma:detconditionmdsl}, we have that the determinant on RHS is $0$ as $V_{A_{1}^{'}}^{'} \cap ... \cap V_{A_{k-1}^{'}}^{'} \neq \B 0$. So, the determinant on LHS is also $0$ which implies $V_{A_1} \cap ... \cap V_{A_k} \neq \B 0$ (again by invoking Lemma \ref{lemma:detconditionmdsl}), which is a contradiction to our assumption. This completes the proof.
\end{IEEEproof}

\section{Proof of Lemma \ref{lemma:removingrowvand}}
\begin{IEEEproof}
\label{proof:removingrowvand}
    Suppose $V$ is $(n,k+1)$-MDS($\ell$) and $V'$ is not $(n,k)$-MDS($\ell$). Then by Lemma \ref{lemma:nkmdsl}, there exists an $(n,k,\ell)$-generic collection $(A_{1}^{'},..,A_{\ell}^{'})$ with $|A_{1}^{'}|+|A_{2}^{'}|+...|A_{\ell}^{'}| = (\ell-1)k$ such that $V_{A_{1}^{'}}^{'} \cap ... \cap V_{A_{\ell}^{'}}^{'} \neq \B 0$. Observe that this means that $|A'_i|\geq 1, \forall i$. Now, we construct a $(n,k+1,\ell)$-generic collection $(A_1,...A_\ell)$ from $(A_{1}^{'},..,A_{\ell}^{'})$ such that $|A_1|+|A_2|+...|A_{\ell}| = (\ell-1)(k+1)$.
    Let $p \in [n]$ and $p \notin (A_1^{'}\cup .. \cup A_{\ell}^{'})$. Such a $p$ exists as $n\geq (\ell-1)(k)+1$. Consider $A_{i}=A_{i}^{'}\cup \{p\}, \forall i\in [\ell-1]$ and $A_{\ell} = A_{\ell}^{'}$. As $(A_{1}^{'},..,A_{\ell}^{'})$ is a $(n,k,\ell)$-generic collection,from definition (\ref{defn:nklgenericcollection}), for any partition $P_1 \cup ...\cup P_s = [\ell]$, we have,
        \begin{align}
    \label{eqn1030}
    \sum_{i=1}^{s}\left |\bigcap_{j \in P_i} A_{j}^{'} \right | \leq (s-1)k.
        \end{align}
    Observe that $|A_i|=|A'_i|+1$, forall $i\in[\ell -1]$, while $|A_\ell|=|A'_\ell|$. Further $p\in \cap_{i=1}^{\ell-1} A_{i}\setminus A_{\ell}$. Thus, $\left|\bigcap_{j \in P_i} A_{j} \right |=\left|\bigcap_{j \in P_i} A'_{j} \right |+1$ if $\ell\notin P_i$, while $\left|\bigcap_{j \in P_i} A_{j} \right |=\left|\bigcap_{j \in P_i} A'_{j} \right |$ for that $P_i$ such that $\ell \in P_i$.
    
    Thus we get
    \begin{align}
        \nonumber
        \sum_{i=1}^{s}\left |\bigcap_{j \in P_i} A_{j} \right | &\leq (s-1)k+(s-1)\\
        \label{eq:nk+1kgenericcollection}
        &\leq (s-1)(k+1)
    \end{align}
    From (\ref{eq:nk+1kgenericcollection}) we can say that $A_1,..A_{\ell}$ form a $(n,k+1,\ell)$-generic collection. WLOG, let $A_\ell=\{1,\hdots,k+1-\delta_{\ell}\}$, for $\delta_{\ell}=(k+1-|A_\ell|)\in\{0,\hdots,k\}$ (as $|A_\ell|=|A'_\ell|\geq 1$) and $\delta_\ell=k+1-|A_\ell|=\delta_{\ell}^{'}+1$.
    From Lemma \ref{lemma:detconditionmdsl}, we have $V_{A_1} \cap ... \cap V_{A_{\ell}} = \B 0$ if and only if
    
    \begin{align}
    \label{eq:detforgenericintersection2}
    \det\begin{pmatrix}[0.75]
    \pi_{A_{1}}^{\delta_1}(\beta_1) & \pi_{A_{2}}^{\delta_2}(\beta_1) & .. & \pi_{A_{\ell-1}}^{\delta_{\ell-1}}(\beta_1)\\
    \pi_{A_{1}}^{\delta_1}(\beta_2) & \pi_{A_{2}}^{\delta_2}(\beta_2) & .. & \pi_{A_{\ell-1}}^{\delta_{\ell-1}}(\beta_2)\\
    . & . & .. & . \\
    . & . & .. & . \\
    \pi_{A_{1}}^{\delta_1}(\beta_{k+1-\delta_{\ell}}) & \pi_{A_{2}}^{\delta_2}(\beta_{k+1-\delta_{\ell}}) & .. & \pi_{A_{\ell-1}}^{\delta_{\ell-1}}(\beta_{k+1-\delta_{\ell}})
    \end{pmatrix}
    \neq 0,
    \end{align}
    
    We now show that the determinant in (\ref{eq:detforgenericintersection2}) is zero so that $V_{A_1} \cap ... \cap V_{A_\ell} \neq \B 0$, leading to a contradiction with the given fact that $V$ is $(n,k+1)$-MDS($\ell$).
    As $p\in (A_{1}\cap A_{2}\cap .. \cap A_{\ell-1})$, the equation (\ref{eqn1052}) holds. 
    \begin{gather}
    \det\begin{pmatrix}
    \pi_{A_{1}}^{\delta_1}(\beta_1) & \pi_{A_{2}}^{\delta_2}(\beta_1) & .. & \pi_{A_{\ell-1}}^{\delta_{\ell-1}}(\beta_1)\\
    \pi_{A_{1}}^{\delta_1}(\beta_2) & \pi_{A_{2}}^{\delta_2}(\beta_2) & .. & \pi_{A_{\ell-1}}^{\delta_{\ell-1}}(\beta_2)\\
    . & . & .. & . \\
    . & . & .. & . \\
    \pi_{A_{1}}^{\delta_1}(\beta_{k+1-\delta_{\ell}}) & \pi_{A_{2}}^{\delta_2}(\beta_{k+1-\delta_{\ell}}) & .. & \pi_{A_{\ell-1}}^{\delta_{\ell-1}}(\beta_{k+1-\delta_\ell})
    \end{pmatrix}\hspace{5cm}\nonumber \\
    \hspace{3cm}= 
    \prod_{j=1}^{k+1-\delta_{\ell}}(\beta_j - \beta_p)\times
    \det\begin{pmatrix}
    \pi_{A_{1}^{'}}^{\delta_{1}^{'}}(\beta_1) & \pi_{A_{2}^{'}}^{\delta_{2}^{'}}(\beta_1) & .. & \pi_{A_{\ell-1}^{'}}^{\delta_{\ell-1}^{'}}(\beta_1)\\
    \pi_{A_{1}^{'}}^{\delta_{1}^{'}}(\beta_2) & \pi_{A_{2}^{'}}^{\delta_{2}^{'}}(\beta_2) & .. & \pi_{A_{\ell-1}^{'}}^{\delta_{\ell-1}^{'}}(\beta_2)\\
    . & . & .. & . \\
    . & . & .. & . \\
    \pi_{A_{1}^{'}}^{\delta_{1}^{'}}(\beta_{k-\delta_{\ell}^{'}}) & \pi_{A_{2}^{'}}^{\delta_{2}^{'}}(\beta_{k-\delta_{\ell}^{'}}) & .. & \pi_{A_{\ell-1}^{'}}^{\delta_{\ell-1}^{'}}(\beta_{k-\delta_{\ell}^{'}})
    \end{pmatrix}.\label{eqn1052}
        \end{gather}
Since $\beta_j\neq \beta_p, \forall j\in A_k$, the first term of RHS of (\ref{eqn1052}) is non-zero. From Lemma \ref{lemma:detconditionmdsl}, We have that the determinant on RHS is $0$ as $V_{A_{1}^{'}}^{'} \cap ... \cap V_{A_{\ell}^{'}}^{'} \neq \B 0$. So, the determinant on LHS is also $0$ which implies $V_{A_1} \cap ... \cap V_{A_{\ell}} \neq \B 0$ (again using Lemma \ref{lemma:detconditionmdsl}), which is a contradiction to our assumption.
\end{IEEEproof}

\section{Proof of Lemma \ref{lemma:sortofshortening}}
\begin{IEEEproof}
\label{proof:sortofshortening}
    WLOG, we assume that the column being left out is $j=n$. Suppose $V$ is an $(n,k)$-MDS($k$) code but $V'$ is not $(n-1,k-1)$-MDS($k-1$). Then there exists some $(n-1,k-1,k-1)$-generic collection $A_{1}^{'},..,A_{k-1}^{'} \subseteq [n-1]$, with 
    \begin{equation}
    \label{eq:removingroweq1}
        |A_{1}^{'}|+|A_{2}^{'}|+..+|A_{k-1}^{'}| = (k-2)(k-1),
    \end{equation}
    such that $V'_{A_{1}^{'}} \cap \hdots \cap V'_{A_{k-1}^{'}} = \B 0$.
    Clearly, there exists an element $p \in [n] \setminus (A_{1}^{'}\cup ..\cup A_{k-1}^{'})$, as we have $A_{i}^{'}\subset [n-1], \forall i\in\{1,\hdots,k-1\}$. Now, using the arguments identical to the proof of Lemma \ref{lemma:removingrow} using this element $p$, we can construct a $(n,k,k)$-generic collection $A_{1},..,A_{k} \subseteq [n]$ with
     \begin{equation}
    \label{eq:removingroweq2}
        |A_1|+|A_2|+..+|A_k| = (k-1)k,
    \end{equation}
    such that such that $V_{A_{1}} \cap \hdots \cap V_{A_{k-1}} = \B 0$. By Lemma \ref{lemma:detconditionmdsl}, this leads to a contradiction, as we are given that $V$ is $(n,k)$-MDS($k$). This completes the proof.

\end{IEEEproof}

    \begin{align}
    \nonumber
    \sum_{j'=\lceil\Delta/\ell\rceil}^{\min(\Delta,n)} \sum_{i=1}^{\min(j,j')} \binom{j}{i}&\binom{n-j}{j'-i}\times \binom{j'}{\leq k}^\ell \nonumber\\
    &\leq \sum_{j'=\lceil\Delta/\ell\rceil}^{\min(\Delta,n)} \binom{n}{j'} \times \binom{j'}{\leq k}^\ell\nonumber\\
    \nonumber
    &\leq   \sum_{j'=\lceil\Delta/\ell\rceil}^{\min(\Delta,n)} \binom{n}{j'}\times \binom{\min(\Delta,n)}{\leq k}^\ell \\
    \label{eq:fieldsizebound}
    &\stackrel{(a)}{\leq}\min\left(2^n,\Delta \left(\frac{en}{\Delta/\ell}\right)^{\Delta}\right)\times \min\left(2^{\ell\min(\Delta,n)},k^\ell\min(\Delta,n)^{k\ell}\right),
    \end{align}
    where in $(a)$ we have used the well known relationships $\sum_{r'=r_1}^{r_2}\binom{m}{r'}\leq 2^m$ and $\binom{m}{r_1}\leq \left(\frac{e\cdot m}{r_1}\right)^{r_1}$, for all non-negative integers $r_1,r_2,m$ such that $r_1,r_2\leq m$.

\section{Proof of Claim \ref{claim:vdetdegree}}
\begin{IEEEproof}
\label{proof:vdetdegree}
As $V$ matrix is obtained using Vandermonde construction, each $V_{A_i}$ ($|A_i| \leq k$) in (\ref{eq:vtodetcondition}) will be of the form 

$$V_{A_i} = \begin{pmatrix}
1 & \hdots &\hdots & 1\\
\beta_{i_1} & . &  . & \beta_{i_{|A_i|}}\\
\beta_{i_1}^{2} & \hdots & . & \beta_{i_{|A_i|}}^{2}\\
. & . & . & . \\
\beta_{i_1}^{k-1} &\hdots & \hdots & \beta_{i_{|A_i|}}^{k-1}
\end{pmatrix},$$
where $A_i=\{i_1,\hdots,i_{|A_i|}\}.$
If the matrix $V_{A_i}$ is square (i.e., $|A_i|=k$), then the degree of the determinant of the above matrix as a polynomial in $\beta_1,..,\beta_n$ is at most $k(k-1)/2$. In the determinant for the matrix in (\ref{eq:vtodetcondition}), 
$\ell$ such submatrices ($V_{A_i}$s) are involved. So, the total degree of the determinant as a polynomial in $\beta_1,..,\beta_n$ is at most $\ell k(k-1)/2 \leq  \ell k^2$.
\end{IEEEproof}

\section{Proof of Claim \ref{claim:noofAcond}}
\begin{IEEEproof}
\label{proof:noofAcond}
    Assume that there are $j$ distinct entries in $\underline{A}$. Note that $j\leq \Delta\triangleq (\ell-1)k$. For some candidate $\underline{A}'\in\mathfrak{A}$, let $j'$ denote the number of distinct entries in $\underline{A}'$. We will use $i$ to denote the number of entries in $\underline{A}'$ which also appear in the set of distinct entries of $\underline{A}$. Note that since we want to count the number of tuples $\underline{A}'$ which are dependent on $\underline{A}$, we have that $1\leq i\leq \min(j,j')$.
    
    Using these observations, we see that the quantity $|\{\underline{A}'\in\mathfrak{A}\colon \underline{A}\not\perp\!\!\!\perp\underline{A}'\}|$ will then be bounded as follows. 
    \begin{equation*}
    |\{\underline{A}'\in\mathfrak{A}\colon \underline{A}\not\perp\!\!\!\perp\underline{A}'\}|\stackrel{(a)}{\leq} \sum_{j'=\lceil\Delta/\ell\rceil}^{\min(\Delta,n)} \sum_{i=1}^{\min(j,j')} \binom{j}{i}\binom{n-j}{j'-i} \times \binom{j'}{\leq k}^\ell
    \end{equation*}
where the term $\binom{j'}{\leq k}\triangleq \sum_{k'=1}^{k}\binom{j'}{k'}$.  Now, we explain the occurrence of the terms on the RHS as follows.
\begin{itemize}
    \item The second summation accounts for the number of possible common entries between $\underline{A}$ and $\underline{A'}$, while the upper-limit in the first summation is because we can have at the most $\min(\Delta,n)$ distinct entries in $\underline{A}'$. The lower-limit in the first summation is because of the condition that $\sum_{j\in[\ell]}|A'_j|=\Delta$ which implies that there should be at least one $A'_j$ with size $\lceil\Delta/\ell\rceil$.
    \item Within the summations, the term $\binom{j}{i}$ counts the number of possible subsets of $i$ distinct entries of $\underline{A}$ that occur in $\underline{A}'$ also.
    \item The term $\binom{n-j}{j'-i}$ counts the number of possible ways to select the remaining $j'-i$ entries in $\underline{A}'$ from the $n-j$ entries in $[n]$ not appearing in $\underline{A}$.
    \item $\binom{j'}{\leq k}^\ell$ occurs to account for the number of ways to obtain the collection $\underline{A}'$ from the specific $j'$ distinct entries that have been chosen from previous steps. 
\end{itemize}
Observing that $\lceil\Delta/\ell\rceil\leq j\leq \min(\Delta,n)$ and $i\geq 1$, we now refine the bound as shown in (\ref{eq:fieldsizebound}), thus completing the proof.
\end{IEEEproof}

\end{appendices}

\end{document}